# TUNGSTEN ISOTOPIC COMPOSITIONS IN STARDUST SiC GRAINS FROM THE MURCHISON METEORITE: CONSTRAINTS ON THE *s*-PROCESS IN THE Hf-Ta-W-Re-Os REGION

Janaína N. Ávila [1,2,3], Maria Lugaro [4], Trevor R. Ireland [1,2], Frank Gyngard [5], Ernst Zinner [5], Sergio Cristallo [6,7], Peter Holden [1], Joelene Buntain [4], Sachiko Amari [5], and Amanda Karakas [8]

[1] Research School of Earth Sciences, The Australian National University, Canberra ACT 0200, Australia; janaina.avila@anu.edu.au

[2] Planetary Science Institute, The Australian National University, Canberra ACT 0200, Australia

[3] Current address: Astronomy Department/IAG, University of São Paulo, São Paulo SP 05508-090, Brazil

[4] Centre for Stellar and Planetary Astrophysics, Monash University, Clayton VIC 3800, Australia

[5] Laboratory for Space Sciences and the Department of Physics, Washington University, One Brookings Drive, St. Louis, MO 63130, USA

[6] Departamento de Física Teórica y del Cosmos, Universidad de Granada, Granada 18071, Spain

[7] Current address: Osservatorio Astronomico di Collurania, INAF, Teramo 64100, Italy

[8] Mount Stromlo Observatory, Australian National University, Weston Creek ACT 2611, Australia

Short Title:

W ISOTOPIC COMPOSITIONS IN STARDUST SiC GRAINS

- 2 -

## ABSTRACT

We report the first tungsten isotopic measurements in stardust silicon carbide (SiC) grains recovered from the Murchison carbonaceous chondrite. The isotopes $^{182,183,184,186}$W and $^{179,180}$Hf were measured on both an aggregate (KJB fraction) and single stardust SiC grains (LS+LU fraction) believed to have condensed in the outflows of low-mass carbon-rich asymptotic giant branch (AGB) stars with close-to-solar metallicity. The SiC aggregate shows small deviations from terrestrial (=solar) composition in the $^{182}$W/$^{184}$W and $^{183}$W/$^{184}$W ratios, with deficits in $^{182}$W and $^{183}$W with respect to $^{184}$W. The $^{186}$W/$^{184}$W ratio, however, shows no apparent deviation from the solar value. Tungsten isotopic measurements in single mainstream stardust SiC grains revealed lower than solar $^{182}$W/$^{184}$W, $^{183}$W/$^{184}$W, and $^{186}$W/$^{184}$W ratios. We have compared the SiC data with theoretical predictions of the evolution of W isotopic ratios in the envelopes of AGB stars. These ratios are affected by the slow neutron-capture process and match the SiC data regarding their $^{182}$W/$^{184}$W, $^{183}$W/$^{184}$W, and $^{179}$Hf/$^{180}$Hf isotopic compositions, although a small adjustment in the *s*-process production of $^{183}$W is needed in order to have a better agreement between the SiC data and model predictions. The models cannot explain the $^{186}$W/$^{184}$W ratios observed in the SiC grains, even when the current $^{185}$W neutron-capture cross section is increased by a factor of two. Further study is required to better assess how model uncertainties (e.g., the formation of the $^{13}$C neutron source, the mass-loss law, the modelling of the third dredge-up, and the efficiency of the $^{22}$Ne neutron source) may affect current *s*-process predictions.

*Subject headings:* dust, extinction — nuclear reactions, nucleosynthesis, abundances — stars: AGB and post-AGB — stars: carbon

# 1. INTRODUCTION

Stardust grains condensed directly from the gas phase present in ancient stellar outflows or stellar ejecta, and thus became part of the interstellar medium from which our Solar System formed about 4.57 Gyr ago. Their stellar origins are indicated by unusual isotopic compositions (essentially for every element) relative to those found in Solar System materials (i.e., terrestrial, lunar, and meteoritic samples). The deviations from the recommended Solar System isotopic abundances, which are taken from compilations of isotopic measurements of terrestrial materials except for H and the noble gases (e.g., Anders & Grevesse 1989; Lodders 2003, 2010), are too large to be explained by mass-fractionation processes or by decay of longer-lived radioactive isotopes. Instead, the observed isotopic compositions can only be explained by nucleosynthetic processes (see reviews by Zinner, 1998, 2004; Clayton and Nittler, 2004; Lodders and Amari, 2005).

Isotopic compositions of several heavy elements have been measured in stardust grains believed to have condensed in the outflows of low-mass carbon-rich asymptotic giant branch (AGB) stars (e.g., Zr, Nicolussi et al. 1997; Mo, Nicolussi et al. 1998a; Sr, Nicolussi et al. 1998b; Ba, Savina et al. 2003; Ru, Savina et al. 2004). The grain data have provided detailed information on the nucleosynthesis of heavy elements produced by the slow neutron capture process (the *s*-process). However, the available data are still scarce or no data exist for some heavy elements. One of the main gaps involves nuclides in the Hf-Ta-W-Re-Os region.

The *s*-process path in the Hf-Ta-W-Re-Os region (Fig. 1) has recently received considerable attention. New experimental data (neutron-capture reaction rates) for Hf, W, and Os nuclides have been reported (Sonnabend et al. 2003; Mohr et al. 2004; Mosconi et al. 2006, 2010a,b; Wisshak et al. 2006; Segawa et al. 2007; Vockenhuber et al. 2007; Marganiec et al. 2009;

72  Fujii et al. 2010), and small anomalies of nucleosynthetic origin in W and Os isotopes have
73  been observed in primitive meteorites (Brandon et al. 2005; Yokoyama et al. 2007, 2011; Qin
74  et al. 2008; Reisberg et al. 2009). Recent *s*-process analyses of the Hf-Ta-W-Re-Os region
75  have identified two major problems. First, it appears that model predictions underestimate the
76  *s*-process contribution to the $^{182}$W solar abundance and, consequently, the $^{182}$W *r*-residual
77  (obtained by subtracting the calculated *s*-process from the observed Solar System abundance)
78  shows a significant positive deviation from the otherwise very smooth rapid neutron-capture
79  process (the *r*-process) solar abundance pattern (see Fig. 12 of Wisshak et al. 2006; and Fig. 6
80  of Vockenhuber et al. 2007). Second, an analysis of the *s*-process flow at the $^{185}$W branching
81  point shows that the predicted $^{186}$Os *s*-process abundance is also somewhat problematic,
82  suggesting a significant overproduction (~ 20%) with respect to its solar abundance
83  (Sonnabend et al. 2003). This is not allowed, theoretically, because $^{186}$Os is an *s*-only isotope,
84  being shielded by the stable $^{186}$W from the chain of radioactive decays that follow the *r*-
85  process. In line with this, isotopic anomalies of nucleosynthetic origin found in Os isotopes
86  measured in primitive meteorites suggest a lower $^{186}$Os/$^{188}$Os *s*-process ratio than current
87  models predict (Brandon et al. 2005; Yokoyama et al. 2007). The problematic $^{182}$W and $^{186}$Os
88  *s*-process abundances and, consequently, the inferred $^{182}$W *r*-residual, may reflect remaining
89  uncertainties related to the experimental cross section data and the current *s*-process models.

90  Tungsten has five stable isotopes: the rare $^{180}$W (0.12% of solar W), which is mainly produced
91  by proton-capture or photodisintegration processes (*p*-process), and $^{182}$W (26.5%), $^{183}$W
92  (14.3%), $^{184}$W (30.6%), and $^{186}$W (28.4%), which are of mixed *s*- and *r*-process origin. As
93  illustrated in Figure 1, the branchings at $^{181,182}$Hf and $^{182,183}$Ta may affect the W *s*-process
94  isotopic pattern, and the branching at $^{185}$W determines the $^{186}$W abundance and may affect the
95  Re and Os isotopic patterns.

Here we report the results of the first W isotopic analyses performed on five large stardust SiC grains (LS+LU fraction) extracted from the Murchison carbonaceous chondrite (Amari et al. 1994). In addition to the individual grains, we also analysed a SiC-enriched bulk sample (KJB fraction, Amari et al. 1994). Carbon-, N-, and Si-isotopic ratios for the KJB fraction and individual SiC grains from the LS+LU fraction have been previously reported by Amari et al. (2000) and Virag et al. (1992), and are reproduced in Figure 2 and Table 1. The mount containing grains from the LS+LU fraction used in this study is the same previously investigated by Virag et al. (1992) and Ireland et al. (1991). Virag et al. (1992) have shown that SiC grains from the LS+LU fraction contain several unique features: some are very large (over 20 μm); many of these large grains appear to have flat and smooth surfaces, unlike the euhedral surfaces observed in smaller grains; and isotopic compositions show clustering for C and Si and even for Ti (Ireland et al. 1991). Recent studies have shown that stardust SiC grains from the LS+LU fraction have interstellar exposure ages ranging from ~ 3 Myr to ~ 1 Gyr (Gyngard et al., 2009a, 2009b; Heck et al., 2009), which implies that the parent stars of the grains must have ended their lives within this time range before the formation of the Solar System. Based on their C-, N-, and Si-isotopic compositions, the five single SiC grains analysed for W are classified as "mainstream grains", and are interpreted to have condensed in the outflows of low-mass (~ 1.5 to 3 $M_\odot$), carbon-rich AGB stars with close-to-solar metallicity (Hoppe et al. 1994; Zinner et al. 2006). The KJB fraction also shows C-, N-, and Si-isotopic signatures consistent with an AGB origin for most of the grains (Amari et al. 2000).

## 2. TUNGSTEN ISOTOPIC MEASUREMENTS

While $^{182}$W and $^{183}$W are free of interferences, the remaining W isotopes have atomic isobaric interferences: $^{180}$W from $^{180}$Ta$^+$ (0.012% of solar Ta) and $^{180}$Hf$^+$ (35.1% of solar Hf), $^{184}$W from $^{184}$Os$^+$ (0.02% of solar Os), and $^{186}$W from $^{186}$Os$^+$ (1.59% of solar Os). Monoxide interferences are also present (e.g. $^{170}$Er$^{16}$O$^+$ and $^{170}$Yb$^{16}$O$^+$ on $^{186}$W$^+$), requiring a mass resolution of m/Δm ~ 8000 for separation. Detailed scans in the mass region of W$^+$ and WO$^+$ isotopic species, sputtered from NIST-610 silicate glass and a SiC ceramic doped with heavy elements, have shown that WO$^+$ species are produced at a higher intensity than W$^+$ species during sputtering by a 10 keV O$_2^-$ primary ion beam (WO$^+$/W$^+$ ~ 3). Furthermore, as previously described by Kinny et al. (1991), under the same analytical conditions YbO$^+$/Yb$^+$ ~ 0.5, and REEO$_2^+$/REEO$^+$ is negligible (< 0.001). Analysing WO$^+$ instead of W$^+$ therefore produces higher yields and minimises interferences from REE$^+$, REEO$^+$, and REEO$_2^+$ species. For these reasons we analysed W isotopes as WO$^+$.

Tungsten isotopic measurements in stardust SiC grains were carried out with a Sensitive High Resolution Ion Microprobe – Reverse Geometry (SHRIMP-RG) at the Australian National University. We performed both "bulk analyses" on an aggregate of many grains from the KJB fraction and "single-grain analyses" on grains from the LS+LU fraction. Nine individual spots on the KJB fraction were analysed for W isotopes. Five out of nine investigated single grains from the LS+LU fraction had sufficiently high W concentrations for isotopic analysis. SHRIMP-RG measurements were performed with an O$_2^-$ primary beam of ~ 5 nA focused to sputter an area of ~ 30 μm in diameter. Before data acquisition, each spot/grain was initially rastered across an area slightly larger than the analytical pit by the beam for ~ 60 s to minimise surface contamination. Secondary ions were extracted at 10 kV and measured by single collector analysis on an ETP$^{TM}$ multiplier in peak-jumping mode.

143   We used two different setups. In the first setup (KJB spots #01 – 08), we measured $^{180}$Hf$^{16}$O$^+$,
144   $^{182}$W$^{16}$O$^+$, $^{183}$W$^{16}$O$^+$, $^{184}$W$^{16}$O$^+$, $^{186}$W$^{16}$O$^+$, $^{188}$Os$^{16}$O$^+$, and $^{189}$Os$^{16}$O$^+$. We monitored OsO$^+$ in order
145   to estimate a potential interference on the $^{186}$W$^{16}$O$^+$ peak. However, no contribution was found
146   in the mass regions of $^{188}$Os$^{16}$O$^+$ and $^{189}$Os$^{16}$O$^+$. In the second setup (KJB spot # 09, and all
147   individual grains), we measured $^{179}$Hf$^{16}$O$^+$, $^{180}$Hf$^{16}$O$^+$, $^{182}$W$^{16}$O$^+$, $^{183}$W$^{16}$O$^+$, $^{184}$W$^{16}$O$^+$, $^{186}$W$^{16}$O$^+$,
148   $^{188}$Os$^{16}$O$^+$, and $^{189}$Os$^{16}$O$^+$. Contributions from $^{180}$W$^{16}$O$^+$ on the $^{180}$Hf$^{16}$O$^+$ peak, $^{180}$Hf$^{18}$O$^+$ on the
149   $^{182}$W$^{16}$O$^+$ peak, and $^{182}$W$^{18}$O$^+$ on the $^{184}$W$^{16}$O$^+$ peak were found to be negligible, so no
150   correction was applied. In addition, we carefully checked the mass region of interest for
151   molecular interferences (Fig. 3) resulting from complex combinations of major elements from
152   the SiC matrix. Low count rates (e.g., ~ 0.4 counts/s at mass 198) were found in the mass
153   region of interest when sputtering a "pure" synthetic SiC. Count rates at mass 198 sputtered
154   from the KJB fraction are usually in the order of 20 counts/s. Individual grains, on the other
155   hand, had count rates between 1.5 and 8 counts/s. The use of a small energy offset (~ 21-24
156   eV) has proved to be quite successful in suppressing complex molecular interferences without
157   significantly compromising the intensity of the atomic species (Ávila et al. 2011, in
158   preparation). In this study, we confirmed this observation for the mass region of interest using
159   a "pure" synthetic SiC reference material. As a result, W (and also Hf) measurements in all
160   individual SiC grains and 5 out of 11 KJB spots (#07 – 11) were carried out on the SHRIMP-
161   RG by combining high-mass resolution with energy filtering. Six KJB spots (#01 – 06) were
162   analysed without energy filtering. We found that all KJB spots have the same W isotopic
163   composition within 2 sigma errors. KJB spots # 10 and 11 were analysed only for $^{179}$Hf/$^{180}$Hf
164   ratios.

165   The acquisition time for each analysis was ~ 6–7 min. Because of the low W abundance in
166   stardust SiC grains, the mass positions of WO$^+$ isotopic species could not be monitored during

167 each individual analysis. Their mass positions were instead maintained from the previous
168 measurement on the standard. We systematically bracketed three unknowns by a suite of
169 standard reference materials (i.e., NIST-610 silicate glass and SiC ceramic). The shifts in
170 peak positions monitored during the analytical sessions were found to be less than 0.002
171 a.m.u. between consecutive standards. The secondary beam was aligned using the QQH
172 monitor, allowing maximum transmission through the source slit. SHRIMP-RG was operated
173 at a mass resolution of m/Δm = 7000 (at 10% peak). The NIST-610 silicate glass and the SiC
174 ceramic doped with heavy elements were used to monitor instrumental mass fractionation
175 (IMF). The data were corrected for an IMF of -12‰ amu$^{-1}$ based on the $^{186}$W/$^{183}$W ratio of
176 1.986 (Jacobsen 2005), and the mean of the $^{186}$W/$^{183}$W measurements of the SiC standards.
177 Mass interferences and background were monitored by periodically analysing a "pure"
178 synthetic SiC and the Au foil (i.e., the substrate on which the grains were deposited). The
179 $^{180}$Hf/$^{184}$W ratios were normalized by applying the relative sensitivity factor (RSF) determined
180 by measuring the SiC ceramic doped with heavy elements.

181 The isotopic ratios obtained with SHRIMP-RG in single-collection mode were calculated
182 using Dodson's time-interpolation algorithm (Dodson 1978) (Table 2), wherein the final
183 isotopic ratios are calculated as means of *N*-1 interpolated ratios (*N* = number of scans). To
184 investigate the hypothesis that the isotopic measurements calculated using this approach could
185 be affected by a systematic positive bias, we also calculated the isotopic ratios from the total
186 counts as suggested by Huss et al. (2011) and Ogliore et al. (2011). We found slight
187 differences (< 5%) in the final ratios, but all are well within the errors. The uncertainties of
188 the W and Hf isotopic ratios measured on stardust SiC grains are dominated by counting
189 statistics, and were calculated from the standard deviation of the sampling distribution (i.e.
190 standard error = S.D./√n). Uncertainties related to the dispersion (i.e. standard deviation) of measurements on the standards during the analytical session were calculated from repeated analysis of NIST-610 and SiC ceramic, and propagated into the uncertainty of each unknown.

# 3. RESULTS

Tungsten and hafnium isotopic compositions are reported in Table 2. The quoted errors are ±1σ. The reference W isotopic composition is taken from Jacobsen (2005). This composition is taken as a terrestrial composition that should also reflect the Solar System composition (e.g., Lodders 2010), which is the case for most refractory elements. Isotope compositions are either reported directly or as delta (δ) values, defined as $\delta R = [(R_{measured}/R_{solar} − 1) \times 1000]$, where the measured isotopic ratios ($R_{measured}$) are expressed as deviations from the reference terrestrial (=solar) isotopic composition ($R_{solar}$) in parts per thousand (‰). Other interelement isotopic ratios are taken from Lodders (2010).

The weighted mean W isotopic ratios obtained for the SiC-enriched bulk sample (KJB fraction), based on 9 measurements (Table 2 and Fig. 4), are $^{182}W/^{184}W = 0.809 \pm 0.014$ ($\delta^{182}W/^{184}W = −64 \pm 17$ ‰), $^{183}W/^{184}W = 0.411 \pm 0.009$ ($\delta^{183}W/^{184}W = −120 \pm 22$ ‰), and $^{186}W/^{184}W = 0.944 \pm 0.024$ ($\delta^{186}W/^{184}W = 18 \pm 26$ ‰). The KJB fraction shows a small deviation from solar composition in both $^{182}W/^{184}W$ and $^{183}W/^{184}W$ ratios, with deficits in $^{182}W$ and $^{183}W$ with respect to $^{184}W$. The $^{186}W/^{184}W$ ratio, however, shows no apparent deviation from the solar value, which raises the possibility of contamination with solar material. Although contamination is possible, we believe it is unlikely since the other ratios (e.g., $^{182}W/^{184}W$ and $^{183}W/^{184}W$) are statistically anomalous.

212   Within the ±1σ uncertainties, the W isotopic ratios determined for two out of five single

213   grains (LU-34 and LU-32) show no deviation from solar composition. On the other hand, two

214   other grains (LU-36 and LU-41) exhibit deficits in $^{182}$W, $^{183}$W, and $^{186}$W with respect to $^{184}$W.

215   Grain LU-20 shows a deviation from the solar value only in the $^{186}$W/$^{184}$W ratio, with both

216   $^{182}$W/$^{184}$W and $^{183}$W/$^{184}$W ratios showing no apparent deviation from solar within errors. Only

217   grain LU-41 has deviations from the solar isotopic ratios that are larger than 2σ. No obvious

218   correlations between W isotopic compositions and C-, N-, and Si-isotopic compositions were

219   found.

220   Additionally, we provide information on the $^{179}$Hf/$^{180}$Hf ratios (Table 2). The weighted mean

221   $^{179}$Hf/$^{180}$Hf of the SiC-enriched bulk sample (KJB fraction), based on 3 measurements, is

222   0.326 ± 0.033 (δ$^{179}$Hf/$^{180}$Hf = −160 ± 85 ‰). This value deviates only slightly from that

223   reported for the Solar System ($^{179}$Hf/$^{180}$Hf = 0.388; Lodders 2010). Three out of five

224   investigated single grains had sufficiently high Hf concentrations for isotopic analysis. Two

225   of them show no apparent deviation from the solar value (within 1σ error). Grain LU-41,

226   which has the largest deficit in $^{182}$W, $^{183}$W, and $^{186}$W with respect to $^{184}$W, also has a significant

227   deficit in $^{179}$Hf with respect to $^{180}$Hf (δ$^{179}$Hf/$^{180}$Hf = −637 ± 121 ‰). All single grains and KJB

228   spot analyses show lower $^{180}$Hf/$^{184}$W ratios than current predictions from *s*-process

229   nucleosynthesis calculations (see section 4).

230   We shall now consider some factors that may affect the present determinations. First, there is

231   a potential problem of contamination with foreign materials and minerals, originating either

232   from the meteorite itself or as a result of sample preparation. Tungsten, unfortunately, may be

233   a contaminant from the sample preparation. Sodium polytungstate was used for density

234   separation of graphite from SiC (Amari et al. 1994). The density separation was followed by

235   several washing procedures in order to remove any contaminants. Based on the behaviour of

236 the isotopic ratios as a function of acquisition time, there is no evidence from our analyses to
237 suggest that the W isotopic compositions are not intrinsic to the grains or result from surface
238 contamination. Furthermore, we did not detect W in the Au-foil, onto which both the KJB and
239 LS+LU fractions had been deposited. Nonetheless, three out of five single grains show no
240 apparent deviation from the solar isotopic ratios (for both W and Hf), and contamination with
241 solar material cannot be completely ruled out. Kashiv (2004) found very high W abundance in
242 stardust SiC grains (W enrichment factor, relative to solar W/Si ratio, of ~ 300 to 8700), much
243 higher than expected based on AGB stellar models and thermodynamic condensation
244 calculations. In contrast to Kashiv (2004), our W concentration measurements are semi-
245 quantitative only. Nevertheless, we found for the KJB fraction a W enrichment factor ~ 10,
246 and for the single grains from the LS+LU fraction an enrichment factor < 3, which are much
247 lower than the values found by Kashiv (2004). This author also suggested that W
248 contamination could be from heavy metal alloy holders used during analytical procedures.
249 Contamination from the Murchison meteorite is another possibility. In order to minimize any
250 residual surface contamination, all spots and grains analysed were initially rastered across an
251 area slightly larger than the analytical pit by the $O_2^-$ primary beam for ~ 60 s before data
252 acquisition.

253 Another complication encountered during ion microprobe analysis of SiC grains is the lack of
254 suitable SiC standards. The accuracy and precision of ion microprobe measurements is known
255 to be highly dependent on the availability of suitable standards because of the variability of
256 ion emission produced by sputtering of solid geological and cosmochemical materials.
257 Standards that are compositionally and structurally similar to the analytical target, i.e., matrix-
258 matched, are therefore highly desirable. Furthermore, assessment of molecular interferences
259 produced by combinations of isotopes of the major elements is particularly important for the analysis of trace elements where even weak molecular interferences may significantly contribute to the mass peak being measured. Naturally occurring (except for stardust SiC grains) and synthetic SiC samples contain very low abundances of trace elements and are therefore not suitable as standard materials. While the synthesis of large doped SiC crystals in the laboratory is extremely difficult, SiC-based ceramics sintered with desirable amounts of trace elements are relatively easy to produce. In order to determine elemental yields and identify possible molecular interferences in the mass region of W isotopes, we prepared a SiC ceramic doped with several trace elements at nominal concentrations between 1 and 3000 ppm. The SiC ceramic has proved to be a useful standard for elemental and isotopic measurements in stardust SiC grains, especially concerning identification of molecular interferences.

## 4. W AND Hf CONDENSATION INTO SiC GRAINS FROM AGB STARS

The $^{180}$Hf/$^{184}$W ratios obtained here for stardust SiC grains (0.041 to 0.276) are lower than *s*-process model predictions for the envelope compositions of carbon-rich, low-mass AGB stars with close-to-solar metallicity (~ 1.42 [1]; Solar System ~ 1.30, Lodders 2010). Clearly, it is not possible to attribute the observed range to nucleosynthetic effects or to contamination with solar material. Hence, we turn to discussing the condensation behaviour of W and Hf to see if this could provide an explanation. According to Lodders & Fegley (1995), under the same conditions of pressure and C/O ratio, W condenses as WC at a temperature ~ 100 K higher than HfC. Both WC and HfC are more refractory than SiC. Thus, one would expect both elements to fully condense into SiC, or at least in the same proportion. Based on our data, this appears not to be the case. One way to explain small fractionations between Hf and W is by having W condense as a metal instead of WC. This can occur if condensation takes place at pressures higher than $10^{-4}$ bars and if W in the circumstellar envelope reaches a concentration of at least 10 times the abundance found in the Solar System (Lodders & Fegley 1995). We note that the W enrichment factor determined for the KJB fraction is ~ 10, which is comparable to the results obtained for other heavy *s*-process elements. However, the single SiC grains from the LS+LU fraction show an enrichment factor < 3. Large fractionation between elements present in SiC relative to the source composition had been previously observed (e.g., Al/Mg, Sr/Ba, Amari et al. 1995; Ni/Fe, Marhas et al. 2008), but there is still no satisfactory explanation for these elemental fractionations.

---

[1] Arithmetic average composition found in the stellar envelope after the last thermal pulse with third dredge-up of the 1.5, 2, 2.5, and 3 $M_\odot$ FRANEC AGB models at Z = 0.01, 0.014, and 0.02 (FRUITY database, http://fruity.oa-teramo.inaf.it:8080/modelli.pl, Cristallo et al. 2011) and 1.25, 1.8, 3, and 4 $M_\odot$ MONASH AGB models at Z = 0.01 and 0.02.

## 5. THE s-PROCESS PATH IN THE W MASS REGION

The *s*-process path in the mass region around W is shown in Figure 1. The *s*-process takes place in the deep He-rich layers of low- and intermediate-mass stars (M ~ 0.8–8 $M_\odot$) during their asymptotic giant branch (AGB) phase of evolution (e.g., Gallino et al. 1998; Busso et al. 1999; Herwig 2005; Zinner et al. 2006; Cristallo et al. 2009). In low-mass AGB stars of close-to-solar metallicity, believed to be the site of origin of mainstream SiC grains, the $^{13}$C($\alpha$, $n$)$^{16}$O reaction is the main neutron source responsible for the production of the bulk of the *s*-process AGB yields. It usually operates under radiative conditions at relatively low temperatures (T ~ 0.9 × $10^8$ K, corresponding to a thermal energy of kT ~ 8 keV), during the interval between episodic He burning (thermal pulses), and results in low neutron densities ($10^6 - 10^7$ neutrons cm$^{-3}$) (Straniero et al. 1995). A second neutron source, the $^{22}$Ne($\alpha$, $n$)$^{25}$Mg reaction, is marginally activated during thermal pulses when the maximum temperature at the bottom of the He-burning shell reaches T ~ 3 × 10$^8$ K (kT ~ 23 keV), resulting in high neutron densities (up to ~ 10$^{10}$ neutrons cm$^{-3}$ in AGB stars of initial mass 1.5 – 3 M$_\odot$). Although neutron captures during these short episodes accounts only for a few percent of the total exposure, they are essential for adjusting the abundance patterns of the *s*-process branchings.

The production of W in the He-intershell during interpulse and thermal pulse phases is strongly affected by branchings at $^{181}$Hf, $^{182}$Hf, $^{182}$Ta, $^{183}$Ta, and $^{185}$W (see Fig. 1). The competition between neutron capture and $\beta$-decay at these branching points can be expressed by a branching factor ($f_n$), calculated from $f_n = \lambda_n/(\lambda_n + \lambda_\beta)$, where $\lambda_n = N_n v_T \langle\sigma\rangle$, and $\lambda_\beta = \ln2/t_{1/2}$ are the neutron-capture rate and the $\beta$-decay rate, respectively. Here, $N_n$, $v_T$, $\langle\sigma\rangle$, and $t_{1/2}$ are the neutron density, the thermal velocity, the Maxwellian-averaged (n, $\gamma$) cross section, and the half-life, respectively. At the typical conditions of operation of the $^{13}$C($\alpha$, $n$)$^{16}$O neutron source (T ~ 0.9 × 10$^8$ K, $N_n$ ~ 10$^6$ – 10$^7$ neutrons cm$^{-3}$), the $^{181}$Hf, $^{182}$Ta, and $^{183}$Ta branching factors are very small (< 0.2%, 1.5%, and < 0.1%, respectively, see Fig. 5a), indicating that $^{181}$Hf, $^{182}$Ta, and $^{183}$Ta will $\beta$-decay to $^{181}$Ta, $^{182}$W, and $^{183}$W, respectively, rather than capture a neutron. Hence, the *s*-process flow will create $^{182}$W, $^{183}$W, and $^{184}$W via $^{181}$Hf($\beta$, $\nu$) $^{181}$Ta(n, $\gamma$) $^{182}$Ta($\beta$, $\nu$) $^{182}$W(n, $\gamma$) $^{183}$W(n, $\gamma$) $^{184}$W. The branching point at $^{185}$W also shows a small probability (~1%) towards $^{186}$W (Fig. 5a), so that no $^{186}$W will be produced during the activation of the $^{13}$C neutron source.

At higher temperatures and neutron density conditions typically found during the AGB thermal pulses (T up to 3 × 10$^8$ K, $N_n$ up to ~ 10$^{10}$ neutrons cm$^{-3}$), the *s*-process production of $^{182}$W, $^{183}$W, and $^{186}$W is different. Under these conditions, the branching factor at $^{181}$Hf is ~ 10% at $N_n$ = 10$^{10}$ neutrons cm$^{-3}$ (Fig. 5b), marginally feeding $^{182}$Hf. The laboratory half-life of

327  $^{182}$Hf is 8.9 × 10$^6$ yrs (Vockenhuber et al. 2004), but at T ~ 8 keV and 23keV (and electron

328  density ~ 5 × 10$^{26}$ cm$^{-3}$) it drops to ~ 41 × 10$^3$ yrs and 12 yrs, respectively, as a result of

329  increased thermal populations of low-lying nuclear states (Takahashi & Yokoi 1987). Despite

330  the enhancement observed in the $^{182}$Hf *β*-decay rate at stellar conditions, its half-life is still

331  sufficiently long to allow neutron captures to occur. Therefore, $^{182}$Hf captures a neutron rather

332  than decay during both interpulse and thermal pulse phases. Provided that the $^{181}$Hf branching

333  point is open, the *s*-process flow will proceed via the sequence $^{181}$Hf(n, γ) $^{182}$Hf(n, γ) $^{183}$Hf(*β*,

334  *ν*) $^{183}$Ta(*β*, *ν*) $^{183}$W, consequently bypassing $^{182}$W, as well as $^{183}$W if the branching point at

335  $^{183}$Ta is also activated (see below). After the neutron flow ceases, $^{182}$W is marginally produced

336  by the decay of $^{182}$Hf. Low-mass AGB stars yield $^{182}$Hf/$^{180}$Hf ratios of ~ 1.79 × 10$^{-2}$, which

337  means that the effect of the $^{181}$Hf branching point is marginal. Only a very small shift of ~ 1 to

338  3% in the $^{182}$W/$^{184}$W ratio due to $^{182}$Hf decay is expected if both Hf and W fully condense into

339  SiC. This shift is likely undetectable given the low $^{180}$Hf/$^{184}$W ratios determined here, which

340  imply a very low *s*-process abundance of $^{182}$Hf in the studied SiC grains. Hence, the

341  observable radiogenic contribution of $^{182}$Hf to its daughter $^{182}$W in the grains would be much

342  less than 1% and thus too difficult to observe.

343  The *s*-process flow that goes through $^{181}$Hf(*β*, *ν*) $^{181}$Ta(n, γ) $^{182}$Ta encounters two other

344  branching points at $^{182}$Ta and $^{183}$Ta. The $^{182}$Ta branching point has a temperature dependent

345  beta-decay rate and a branching factor > 80% (Fig. 5b) during the high neutron density

346  produced by the $^{22}$Ne source, and may cause the *s*-process flow to partially bypass $^{182}$W. The

347  branching factor at $^{183}$Ta is ~ 50% at T ~ 3 × 10$^8$ K and $N_n$ = 10$^{10}$ neutrons cm$^{-3}$ (Fig. 5b),

348  causing the *s*-process flow to partially bypass $^{183}$W. The branching factor at $^{185}$W is enhanced

349  similarly to the $^{182}$Ta branching factor during thermal pulses (Fig. 5b), feeding $^{186}$W, which

350  results in a smaller *s*-process contribution to both $^{186}$Os and $^{187}$Os, due to the presence of the long-living nuclei $^{187}$Re. The $^{186}$W/$^{184}$W ratios reported here suggest activation of the branching point at $^{185}$W. This result is, in principle, in disagreement with $^{96}$Zr depletions observed in SiC grains that indicate that the $^{22}$Ne neutron source was weak in the parent stars of the grains (Nicolussi et al. 1997; Lugaro et al. 2003). However, at conditions of T ~ 23 keV and N$_n$ > 5 ×10$^8$ neutrons cm$^{-3}$, the branching point at $^{95}$Zr shows a smaller branching factor than $^{185}$W (Fig. 5b). Therefore, at the same conditions, the $^{185}$W(n, γ)$^{186}$W reaction shows a higher probability to occur than the $^{95}$Zr(n, γ)$^{96}$Zr reaction. Information about the Zr isotopic compositions in the same grains measured for W would help to better constrain the physical conditions of the *s*-process in the parent stars of the grains.

From Figure 1, it can be seen that there is only one path that leads from $^{179}$Hf to $^{180}$Hf [2], whose relative abundances are not affected by any branching point. If a steady flow is achieved along the *s*-process path, the local equilibrium approximation applies $\langle\sigma\rangle_{(A)} N_{s(A)} \sim \langle\sigma\rangle_{(A-1)} N_{s(A-1)}$, where $\langle\sigma\rangle_{(A)}$ is the Maxwellian averaged (n, γ) cross section of the isotope A, and $N_s$ its *s*-process abundance (Clayton 1983). From this simple formulation, we can infer that the ratio of the *s*-process contributions to $^{179}$Hf and $^{180}$Hf are approximately equal to the inverse ratios of their neutron-capture cross sections. At thermal energies of ~ 8 and 23 keV, the inverse ratio of the $^{179}$Hf/$^{180}$Hf neutron-capture cross sections yields $\langle\sigma\rangle_{(^{180}Hf)}/\langle\sigma\rangle_{(^{179}Hf)} = 0.150$ and 0.175, respectively, with an uncertainty of ~ 1.5% (Dillmann et al. 2006). The grain showing the most anomalous W and Hf isotopic ratios, LU-41, has $^{179}$Hf/$^{180}$Hf = 0.141 ± 0.047, in very good agreement with the ratio expected from the values of neutron-capture cross sections, thus indicating an almost pure *s*-process signature in this grain.

---

[2] A branching point at $^{179}$Hf may be activated because this stable nucleus becomes unstable in stellar conditions. This effect is insignificant in our context, as the beta-decay half-life of $^{179}$Hf is ~ 30 yrs at T = 3 × 10$^8$ K, however, in some conditions it may lead to the production of $^{180}$Ta (actually an isomeric state, $^{180}$Ta$^m$), the least abundant nucleus in the Solar System (e.g., Käppeler et al. 2004). A weak *s*-process branching may also be activated via neutron captures on $^{179}$Hf, feeding the 8$^-$ isomeric state in $^{180}$Hf ($t_{1/2}$ = 5.5. h), which then decays to $^{180}$Ta$^m$ (Beer & Ward 1981).

## 6. DISCUSSION

Figures 6 and 7 show the W and Hf isotopic ratios of the stardust SiC grains together with *s*-process model predictions for the envelope compositions of low-mass AGB stars. These predictions represent the mixing between two components, one close to the solar composition, representing the initial composition of the stellar envelope (in stardust studies this is traditionally referred to as the N-component), the other with isotopic characteristics close to those predicted for pure *s*-process (the G-component) (e.g., Zinner et al. 1991; Nicolussi et al. 1997; Savina et al. 2004). The magnitude of the departure from the N-component towards the G-component depends on the efficiency and the number of mixing episodes, also known as third dredge-ups (TDUs), which occur when the convective envelope penetrates into the He intershell zone. The TDU causes newly synthesized $^{12}$C and *s*-processed material to be mixed into the convective envelope of the star (e.g., Lugaro et al. 2003).

Two different sets of models (FRANEC, Cristallo et al. 2009, 2011; MONASH, Karakas 2010, Karakas et al. 2010) with a range of stellar masses (1.25 – 4 $M_\odot$) and metallicities (Z = 0.01, 0.014, and 0.02) were investigated. Of relevance here, in the FRANEC code the neutron-capture cross sections for the W isotopes are taken from Bao et al. (2000), and correspond to the values published by Macklin et al. (1983) for $^{182}$W, $^{183}$W, $^{184}$W, and $^{186}$W, and the theoretical value adopted by Bao et al. (2000) for $^{185}$W. In the MONASH code, the neutron-capture cross sections for the W isotopes are taken from the KADoNiS Database (Dillmann et al. 2006), and correspond to the values published by Macklin et al. (1983) for

$^{182}$W and $^{183}$W, Marganiec et al. (2009) for $^{184}$W and $^{186}$W, and Mohr et al. (2004) for $^{185}$W. The models presented here have been selected for C/O > 1 in the envelope so that the condition for the formation of SiC is satisfied.

In spite of all the differences between the two evolutionary codes, which employ a variety of input physics also concerning mixing and reaction rates (see Cristallo et al. 2009, 2011; Karakas 2010; and Karakas et al. 2010 for details), the isotopic compositions predicted by the two sets of models are in fair agreement with one another. Both sets of models present a good match with the SiC data for the $^{182}$W/$^{184}$W, $^{183}$W/$^{184}$W, and $^{179}$Hf/$^{180}$Hf ratios (Fig. 6a, c and Fig. 7a, b, d, e). As already demonstrated by Lugaro et al. (2003) in relation to Sr, Zr, and Ba isotopic ratios, the marginal activation of the $^{22}$Ne neutron source in the thermal pulses is necessary to best reproduce some of the single SiC data. In the case of $^{182}$W/$^{184}$W and $^{183}$W/$^{184}$W ratios, the $^{22}$Ne neutron source allows the activation of the $^{182}$Ta and $^{183}$Ta branching points, lowering the $^{182}$W/$^{184}$W and $^{183}$W/$^{184}$W ratios. This is clearly illustrated by comparison in Figure 7d and 7e of the 1.8 M$_\odot$ model, where the $^{22}$Ne neutron source is not activated, as the temperature in the thermal pulses is always below $2.66 \times 10^8$ K, with the higher mass models. The SiC grains that are isotopically solar within errors could be explained either as contamination with solar material, or as originating in a very low-mass star (the 1.25 M$_\odot$ model) where TDU episodes can only slightly modify the envelope's pristine composition. The KJB average measurement is also well explained as the average of different stellar masses, except that the $^{183}$W/$^{184}$W ratio is below the average of the stellar models (Fig. 7b, e). Also, the $^{183}$W/$^{184}$W ratio observed in the most extreme grain, LU-41, is slightly lower than the lowest predicted value (but note that the error bars are 1$\sigma$).

On the other hand, most models predict lower $^{186}$W/$^{184}$W ratios than those measured in the grains (Fig. 6b, d and Fig. 7c, f). The model predictions move towards higher $^{186}$W/$^{184}$W ratios with increasing number of thermal pulses, particularly for models of higher masses and lower metallicities, as the $^{22}$Ne neutron source and thus the $^{185}$W branching point are more activated in these models. However, they fail to reach the observed values. While some ad hoc solutions may be found for the single grain data (e.g., the 4 M$_\odot$ model reaches the $^{186}$W/$^{184}$W ratio measured in grain LU-41 and almost also the value observed in grain LU-36, when considering that the error bars are 1 sigma), the KJB ratio is particularly puzzling; it agrees with the solar value within a relatively small error while it should represent the average of different stellar sources with different masses and metallicities, or, in case of contamination, should sit on the mixing line connecting the models and the solar composition point.

A question that needs to be addressed is whether or not a revision of the neutron-capture cross sections would improve the match between model predictions and the SiC data. For $^{182}$W and $^{183}$W, the uncertainties in the recommended neutron-capture cross sections are very small (~ 3%), but the measurements are quite old (Macklin et al. 1983) and no revision has been obtained since. The possibility that the $^{182}$W neutron-capture cross section had been overestimated by 20%-30% in the past was recently suggested by Wisshak et al. (2006) and Vockenhuber et al. (2007) based on the observation that the $^{182}$W $r$-residual shows a significant positive deviation from the otherwise very smooth $r$-process solar abundance pattern (see Fig. 12 of Wisshak et al. 2006; or Fig. 6 of Vockenhuber et al. 2007). Also, a smaller $^{182}$Ta neutron-capture cross section, which is based on theoretical calculations with a relatively large uncertainty (~ 16%), could lead to a lower $^{182}$Ta branching factor, resulting in a higher $s$-process contribution to $^{182}$W and a lower $^{182}$W $r$-residual (see Table 13 of Vockenhuber et al. 2007). However, any major changes resulting in a higher $^{182}$W $s$-process production would lead to higher $^{182}$W/$^{184}$W ratios and a mismatch with the SiC data reported here. For example, a 30% reduction of the $^{182}$W cross section in the MONASH 3 M$_\odot$ and Z =

0.01, and 3 $M_\odot$ and Z = 0.02 models results in a ~ 35% increase in the final $^{182}$W/$^{184}$W ratio in the stellar envelope, which is well above the values observed in the SiC grains. On the other hand, our data suggest that a small adjustment (toward lower values) in the *s*-process production of $^{183}$W is needed in order to obtain a better agreement between our SiC data and the *s*-process model predictions. This may be achieved by adopting a higher (~ 30%) $^{183}$W neutron-capture cross section.

As for the unstable $^{185}$W, its neutron-capture cross section has been recently derived from the inverse $^{186}$W(γ, n)$^{185}$W photodisintegration reaction (Sonnabend et al. 2003; Mohr et al. 2004) with an uncertainty between 10 and 15% and a difference between the two measurements of 24% at 30 keV. These values lead to an overproduction of the solar *s*-only $^{186}$Os of at least 20% (Sonnabend et al. 2003). Several hypotheses have been raised to resolve this puzzle. Sonnabend et al. (2003) suggested that the stellar model apparently overestimates the *β*-decay part and/or underestimates the neutron-capture part of the $^{185}$W branching point. Meyer and Wang (2007), on the other hand, proposed that this problem could be resolved either by increasing the $^{186}$Os neutron-capture cross section by ~ 20% or by increasing the branching factor at $^{186}$Re. An increase in the $^{186}$Os neutron-capture cross section would also help in explaining the low $^{186}$Os/$^{188}$Os *s*-process ratio derived from isotopic anomalies observed in primitive meteorites (Brandon et al. 2005; Yokoyama et al. 2007). However, a recent measurement of the $^{186}$Os neutron-capture cross section by Mosconi et al. (2010a) excludes this hypothesis. As noted by Reisberg et al. (2009), the mismatch between the inferred $^{186}$Os/$^{188}$Os *s*-process ratio, derived from isotopic anomalies observed in primitive meteorites (Brandon et al. 2005; Yokoyama et al. 2007), and model predictions could be resolved by using a recent measurement of the $^{188}$Os neutron-capture cross section (Mosconi et al. 2007), which is ~ 27% lower than the recommended value of Bao et al. (2000). Sonnabend et al.

(2003) suggested an increase of the $^{185}$W neutron-capture cross section (by about 60%), which would lead to a decrease of $^{186}$Os and an increase of the $^{186}$W abundance. We computed AGB models using the recommended $^{185}$W neutron-capture cross section multiplied by a factor of 2 to see if this very large change would help in matching the $^{186}$W/$^{184}$W grain data. We obtain an increase of ~ 20% in the final $^{186}$W/$^{184}$W ratio in the stellar envelope of the MONASH 3 M$_\odot$ and Z = 0.01, and 3 M$_\odot$ and Z = 0.02 models, but it still lower than the observed stardust SiC data.

We also checked the impact of the uncertainties in the $\beta$-decay rates of the unstable isotopes $^{181}$Hf, $^{182}$Ta, and $^{185}$W. At a temperature of $3 \times 10^8$ K, Goriely (1999) reported uncertainties of roughly a factor of three for $^{181}$Hf and $^{182}$Ta and of roughly 40% for $^{185}$W. Variations were found to be small, at most 20% in the $^{182}$W/$^{184}$W ratios, mostly due to the uncertainties in the decay rate of $^{182}$Ta. Further, we checked the impact of the uncertainties in the theoretical neutron-capture cross sections of $^{181}$Hf and $^{182}$Ta by considering the theoretical estimates reported in the KADoNiS Database (Dillmann et al. 2006) on the basis of different codes used to compute the rates. These uncertainties do not lead to significant variations on the model predictions, at most 16% in the $^{182}$W/$^{184}$W ratios when the neutron-capture cross section of $^{182}$Ta was varied by a factor of two.

In summary, the SiC data do not support the problem of the *s*-process underproduction of $^{182}$W that follows from the high solar *r*-process residual. However, it is important to note that the *r*-process residuals are calculated from *s*-process predictions obtained from one stellar model (or at most from the average of two different masses at the same metallicity, e.g., Arlandini et al. 1999) or from phenomenological models (e.g. Käppeler et al. 1982). A more realistic approach would be to derive them from models of the chemical evolution of the Galaxy (e.g., Travaglio et al. 1999). This is because the Solar System composition is the

489   result of nucleosynthesis from many generations of AGB stars of many different masses and
490   metallicities. This more realistic description of the solar *s*-process abundances and of the *r*-
491   process residuals may help solving the discrepancies found in the Hf-Ta-W-Re-Os mass
492   region and needs to be pursued.

493   The SiC data qualitatively support the solution of the $^{186}$Os overproduction problem according
494   to which the $^{185}$W branching point should be activated more strongly. However, while the
495   $^{186}$Os problem is solved by increasing the neutron-capture cross section of $^{185}$W by ~ 60%
496   (Sonnabend et al. 2003), our models do not match the SiC data even if this cross section is
497   increased by a factor of two! Another way to increase the $^{186}$W/$^{184}$W ratio would be to
498   decrease the production of $^{184}$W. The $^{186}$W nuclide is created during thermal pulses, when the
499   $^{185}$W branching point is open. In contrast, $^{184}$W is created within the $^{13}$C pocket during the
500   radiative $^{13}$C burning. Thus, the production of $^{184}$W is directly correlated with the $^{13}$C (and $^{14}$N)
501   abundances within the pocket, which in turn depend on the mixing mechanism that allows a
502   few protons to penetrate from the envelope during the TDU into the underlying $^{12}$C rich
503   intershell. The current uncertainties affecting this mixing are still large, so that a clear
504   description is still not available and the amount of $^{13}$C in the pocket may still be treated as a
505   free parameter. In the FRANEC code, the $^{13}$C abundance within the pocket is derived from the
506   application of an exponentially decreasing profile of convective velocities at the inner border
507   of the convective envelope. In the MONASH code, the $^{13}$C abundance within the pocket is
508   derived from artificial inclusion of an exponentially decreasing abundance of protons in the
509   He intershell. The inclusion of physical mechanisms not explicitly treated in our codes, such
510   as rotation, could modify the $^{13}$C and $^{14}$N abundances within the pocket before and during the
511   activation of the $^{13}$C$(\alpha, n)^{16}$O reaction, thus leading to different $^{186}$W/$^{184}$W ratios. For example,
512   the calculations of Arlandini et al. (1999) used a very different $^{13}$C pocket than ours and 513   produced $^{186}$W/$^{184}$W = 0.66 (together with $^{183}$W/$^{184}$W = 0.35). However, these models were
514   targeted specifically to match the *s*-only solar distribution and are hence not appropriate to be
515   compared to the composition of stardust SiC grains. In order to test the effects of a different
516   $^{13}$C profile in our calculations, we changed in the MONASH 3 M$_\odot$, Z = 0.02 model the extent
517   in mass of the $^{13}$C pocket from 0.002 M$_\odot$ to 0.0005 M$_\odot$. We obtained $^{186}$W/$^{184}$W = 0.58,
518   however the $^{183}$W/$^{184}$W ratio is also closer to the solar value (0.43).

519   Note that although our SiC data allowed us to address some discrepancies found in the solar
520   *s*- and *r*-abundance distributions in the Hf-Ta-W-Re-Os mass region, the stardust grains
521   provide a more direct constraint on the *s*-process in low-mass AGB stars rather than the solar
522   *s*-process component. Furthermore, the calculation of solar *s*- and *r*-abundance distributions
523   suffers from significant uncertainties. Goriely (1999) used an extended parametric *s*-process
524   model, the "multi-event *s*-process" model described in Goriely (1997), to analyse the impact
525   of nuclear and observational uncertainties on the solar *r*-process residual distribution. The
526   final result, presented in Figure 8 of Goriely (1999), clearly shows that the smoothness of the
527   *r*-process residual curve in the Hf-Ta-W-Re-Os region is well reproduced within the
528   uncertainties. This would still be true when considering the new measurements of the Hf
529   neutron-capture cross-sections of Wisshak et al. (2006), which result in a downward shift of
530   the *r*-residual of $^{180}$Hf. There are other uncertainty factors that may also affect the solar *s*- and
531   *r*-abundances predictions, like those arising from galactic and stellar evolution models. The
532   combined effect of all uncertainties precludes any definite conclusion about the solar *s*- and *r*-
533   abundance distributions to be made to the level needed for a detailed comparison with the
534   stardust data.

535

## 7. CONCLUSIONS

536

537    We report the first W isotopic anomalies in mainstream stardust SiC grains recovered from

538    the Murchison carbonaceous chondrite. Comparisons between grain data and *s*-process model

539    predictions for the envelope compositions of low-mass AGB stars show that a single choice

540    of stellar mass and metallicity cannot account for the range of isotopic ratios observed in the

541    SiC grains. There is an overall match between the SiC data and the theoretical predictions for

542    the $^{182}W/^{184}W$ and $^{183}W/^{184}W$ ratios, particularly if a 30% higher neutron-capture cross section

543    is employed for $^{183}W$. However, the AGB models fail to reproduce the high $^{186}W/^{184}W$ ratios

544    observed in the grains. We note, however, that many uncertainties affect stellar evolutionary

545    computations of the *s*-process. Among them are the formation of the $^{13}C$ pocket, the mass-loss

546    law, the TDU, and the efficiency of the $^{22}Ne$ neutron source. For example, the inclusion of

547    rotation and magnetic fields may affect the operation of the $^{13}C$ neutron source, and a lower

548    efficiency of this neutron source may decrease the production of $^{184}W$, thus increasing the

549    $^{183}W/^{184}W$ ratio. Also a milder mass-loss history would lead to a larger number of thermal

550    pulses and, therefore, possibly to a larger $^{186}W/^{184}W$ ratio. Similarly, an increase of the $^{22}Ne(\alpha,$

551    $n)$ cross section could lead to a higher neutron density, thus increasing the $^{186}W$ production,

552    but perhaps worsening the match with the $^{96}Zr/^{94}Zr$ ratio observed in the grains (Lugaro et al.

553    2003). We intend to explore these hypotheses in the future. Finally, further W isotopic

554    measurements with better precision on additional grains would be extremely helpful. They

555    could shed light on the possibility of contamination affecting the current W measurements

556    and confirm the high $^{186}W/^{184}W$ ratios reported here.

557

558

## ACKNOWLEDGEMENTS

J.N. Ávila was supported by a Ph.D. scholarship (Grant#200081/2005-5) of the Brazilian National Council for Scientific and Technological Development (CNPq). She acknowledges the Australian National University through an ANU Vice-Chancellor's Higher Degree Research Travel Grant. M. Lugaro is an ARC Future Fellow and a Monash Research Fellow. T. R. Ireland acknowledges support by ARC grants DP0342772 and DP0666751. E. Zinner acknowledges support by NASA grant NNX08AG71G. S. Cristallo acknowledges support by the Spanish Grant AYA2008-04211-C02-02 and FPA2008-03908 from the MEC. S. Cristallo thanks Roberto Gallino for fruitful scientific discussions. We are grateful to Roy Lewis for providing the LS+LU grains. We thank an anonymous referee for their comments and Eric Feigelson for handling of this manuscript.

FIGURE CAPTIONS

Figure 1: Part of the nuclide chart showing the *s*-process nucleosynthesis path in the region of Hf-Ta-W-Re-Os (modified from Hayakawa et al. 2005; and Dillmann et al. 2006). Percent abundances (non-italic) are shown for each stable isotope (solid boxes) and laboratory half-lives (italic) for each unstable isotope (dashed line boxes). Half-lives at stellar temperatures may be different, as discussed in the text. The main *s*-process path is shown as a bold line and branches and secondary paths are shown as finer lines, *s*-only isotope $^{186}$Os is indicated by a bold box.

Figure 2: Silicon-, C-, and N-isotopic ratios of stardust SiC grains from the LS+LU fraction (white squares, from Virag et al. 1992) analysed in the present study. Data for mainstream SiC grains from previous analyses (Hynes & Gyngard 2009) are shown for comparison (grey squares). Data for the SiC-enriched bulk sample (KJB fraction, black square) from Amari et al. (2000) are also plotted. Error bars from previous measurements are omitted for clarity. Black dashed lines indicate the solar ratios (Lodders 2010) as inferred from terrestrial composition. (a) Si-isotopic ratios expressed as deviations (δ-values) from the reference terrestrial (=solar) isotopic composition in parts per thousand (‰). Error bars are smaller than the symbols. Also shown is the so-called mainstream correlation line indicated by the solid line, with slope of 1.35 (Zinner et al. 2006). (b) $^{14}$N/$^{15}$N ratios plotted against $^{12}$C/$^{13}$C ratios. Error bars are smaller than the symbols.

Figure 3: SHRIMP-RG mass scans of $^{182}$W$^{16}$O$^+$ (a), $^{183}$W$^{16}$O$^+$ (b), $^{184}$W$^{16}$O$^+$ (c), and $^{186}$W$^{16}$O$^+$ (d), obtained in NIST-610 silicate glass, a "pure" synthetic SiC, and a SiC ceramic doped with heavy elements. Energy offset = 0 eV and m/Δm ~ 7000 (at 10% peak).

Figure 4: Tungsten isotopic compositions determined for the SiC-enriched sample (KJB fraction). Each bar represents an individual measurement. Bold and dashed lines indicate solar ratios (as inferred from terrestrial composition, Jacobsen 2005) and SiC weighted mean ratios, respectively. Box heights are 2σ. Weighted means are also reported as deviations (δ-values) from the reference terrestrial (=solar) isotopic composition in parts per thousand (‰). MSWD = mean square weighted deviation.

Figure 5: Branching factor ($f_n$) of $^{181}$Hf, $^{182}$Hf, $^{182}$Ta, $^{183}$Ta, $^{185}$W, and $^{95}$Zr at kT = 8 keV (a) and 23 keV (b) as a function of neutron density. We used the β-decay rates reported by Takahashi & Yokoi (1987) and the latest accepted neutron-capture rates from the KADoNiS Database (http://www.kadonis.org/; Dillmann et al. 2006), except for the $^{183}$Ta, where the neutron-capture rate from JINA REACLIB database (http://groups.nscl.msu.edu/jina/reaclib/db/; Cyburt et al. 2010) was used. All values were calculated for an electron density of $5 \times 10^{26}$ cm$^{-3}$. The branching factor (%) depends on the neutron-capture and $β^-$ decay rates and indicates the probability that an unstable isotope will capture a neutron rather than decay. The grey area in (a) and (b) corresponds to the conditions typically found during interpulse and thermal pulse phases, respectively, in low-mass AGB stars.

Figure 6: (a, c) $^{182}$W/$^{184}$W plotted against $^{183}$W/$^{184}$W and (b, d) $^{186}$W/$^{184}$W plotted against $^{183}$W/$^{184}$W for the SiC-enriched sample (KJB fraction) and single SiC grains (LS+LU fraction). Error bars are 1σ. The grey dashed lines in each plot give the best linear fit through all single SiC grains and the SiC-enriched sample. The black solid lines shown in each plot give the best linear fit through grains LU-36 and LU-41, and the SiC-enriched sample; grains LU-34, LU-32, and LU-20 were not included in the fit because their W isotopic signature may be affected by contamination with solar

material. Black dashed lines indicate the solar ratios (as inferred from terrestrial composition, Jacobsen 2005). SiC data are compared with *s*-process model predictions (FRANEC and MONASH models) for the envelope compositions of low-mass AGB stars of different masses and metallicities (see text for details). Predictions are only plotted when the C/O in the stellar envelope reaches values higher than 1.

Figure 7: (a, d) $^{182}W/^{184}W$ versus $^{179}Hf/^{180}Hf$, (b, e) $^{183}W/^{184}W$ versus $^{179}Hf/^{180}Hf$, and (c, f) $^{186}W/^{184}W$ versus $^{179}Hf/^{180}Hf$ for the SiC-enriched sample (KJB fraction) and single SiC grains (LS+LU fraction). Error bars are 1σ. The gray dashed lines in each plot give the best linear fit through all single SiC grains and the SiC-enriched sample. The black solid lines shown in each plot give the best linear fit through grain LU-41 and the SiC-enriched sample; grains LU-34 and LU-32 were not included in the fit because their W isotopic signature may be affected by contamination with solar material. Black dashed lines indicate the solar ratios (as inferred from terrestrial composition, Jacobsen 2005 and Lodders 2010). SiC data are compared with *s*-process model predictions (FRANEC and MONASH models) for the envelope compositions of low-mass AGB stars of different masses and metallicities (see text for details). Predictions are only shown when the C/O in the stellar envelope reaches values higher than 1.

**Table 1:** C-, N- and Si-isotopic compositions of stardust SiC grains from the KJB and LS+LU fractions measured for W and Hf isotopes. Isotopic ratios are reproduced from Amari et al. (2000) and Virag et al. (1992). Errors are 1σ.

| Grain/ Spot | Size ($\mu$m) | $^{12}$C/$^{13}$C ± 1σ | $^{14}$N/$^{15}$N ± 1σ | $\delta^{29}$Si/$^{28}$Si [a] ± 1σ (‰) | $\delta^{30}$Si/$^{28}$Si [a] ± 1σ (‰) |
|---|---|---|---|---|---|
| KJB fraction (Murchison SiC-enriched sample) | | | | | |
| KJB | 0.49 | 37.0 ± 0.4 | 521 ± 60 | 24.6 ± 1.3 | 37.8 ± 3.4 |
| LS+LU fraction (Murchison single grains) | | | | | |
| LU-34 | 7 x 9 | 54.3 ± 0.4 | 1889 ± 13 | 86.4 ± 3.0 | 73.2 ± 3.4 |
| LU-36 | 23 x 23 | 49.1 ± 0.4 | 408 ± 20 | 37.9 ± 2.5 | 41.1 ± 3.1 |
| LU-32 | 5 x 13 | 63.0 ± 0.4 | 1088 ± 14 | 55.3 ± 2.5 | 47.8 ± 3.2 |
| LU-20 | 15 x 26 | 48.6 ± 0.4 | 935 ± 22 | 38.6 ± 2.6 | 45.0 ± 3.5 |
| LU-41 | 13 x 13 | 48.4 ± 0.3 | 678 ± 15 | 43.3 ± 3.4 | 44.3 ± 2.9 |

[a] $\delta^{i}$Si/$^{28}$Si (‰) = [($^{i}$Si/$^{28}$Si)$_{measured}$/($^{i}$Si/$^{28}$Si)$_{solar}$ -1] × 10$^{3}$.

**Table 2:** Tungsten and hafnium isotopic ratios determined in stardust SiC grains. Errors are 1σ.

| Spot/ Grain | $^{182}W/^{184}W$ ± 1σ | $\delta^{182}W/^{184}W$ [b] ± 1σ (‰) | $^{183}W/^{184}W$ ± 1σ | $\delta^{183}W/^{184}W$ [b] ± 1σ (‰) | $^{186}W/^{184}W$ ± 1σ | $\delta^{186}W/^{184}W$ [b] ± 1σ (‰) | $^{179}Hf/^{180}Hf$ ± 1σ | $\delta^{179}Hf/^{180}Hf$ [c] ± 1σ (‰) | $^{180}Hf/^{184}W$ [d] ± 1σ |
|---|---|---|---|---|---|---|---|---|---|
| Terrestrial [a] | 0.865 | | 0.467 | | 0.928 | | 0.388 | | 1.302 |
| *KJB fraction (Murchison SiC-enriched sample)* | | | | | | | | | |
| KJB-01 | 0.725 ± 0.054 | −162 ± 63 | 0.395 ± 0.027 | −154 ± 57 | 0.755 ± 0.077 | −186 ± 83 | n.a. | | n.a. |
| KJB-02 | 0.824 ± 0.044 | −48 ± 51 | 0.434 ± 0.050 | −70 ± 108 | 0.858 ± 0.052 | −76 ± 56 | n.a. | | 0.095 ± 0.013 |
| KJB-03 | 0.817 ± 0.024 | −56 ± 28 | 0.399 ± 0.015 | −147 ± 31 | 0.985 ± 0.029 | 62 ± 31 | n.a. | | 0.046 ± 0.002 |
| KJB-04 | 0.752 ± 0.045 | −130 ± 52 | 0.395 ± 0.029 | −154 ± 62 | 0.904 ± 0.056 | −25 ± 60 | n.a. | | 0.041 ± 0.004 |
| KJB-05 | 0.825 ± 0.043 | −46 ± 50 | 0.394 ± 0.026 | −157 ± 57 | 0.954 ± 0.054 | 28 ± 58 | n.a. | | 0.097 ± 0.015 |
| KJB-06 | 0.838 ± 0.056 | −31 ± 64 | 0.461 ± 0.037 | −12 ± 78 | 0.963 ± 0.044 | 39 ± 47 | n.a. | | 0.059 ± 0.009 |
| KJB-07 | 0.845 ± 0.055 | −23 ± 63 | 0.454 ± 0.036 | −28 ± 76 | 0.995 ± 0.065 | 72 ± 70 | n.a. | | 0.057 ± 0.012 |
| KJB-08 | 0.829 ± 0.058 | −42 ± 67 | 0.451 ± 0.038 | −35 ± 80 | 0.996 ± 0.068 | 74 ± 72 | n.a. | | 0.068 ± 0.007 |
| KJB-09 | 0.817 ± 0.056 | −55 ± 64 | 0.436 ± 0.036 | −66 ± 77 | 0.902 ± 0.064 | −28 ± 69 | 0.338 ± 0.060 | −130 ± 155 | 0.069 ± 0.007 |
| KJB-10 | n.a. | | n.a. | | n.a. | | 0.318 ± 0.059 | −181 ± 153 | n.a. |
| KJB-11 | n.a. | | n.a. | | n.a. | | 0.322 ± 0.056 | −170 ± 144 | n.a. |
| *Weighted average* | 0.809 ± 0.014 | −64 ± 17 | 0.411 ± 0.009 | −120 ± 22 | 0.944 ± 0.024 | 18 ± 26 | 0.326 ± 0.033 | −160 ± 85 | 0.050 ± 0.005 |
| *LS+LU fraction (Murchison single grains)* | | | | | | | | | |
| LU-34 | 0.866 ± 0.067 | 1 ± 77 | 0.485 ± 0.044 | 38 ± 95 | 0.887 ± 0.069 | −44 ± 75 | 0.475 ± 0.076 | 223 ± 195 | 0.194 ± 0.035 |
| LU-36 | 0.722 ± 0.070 | −165 ± 80 | 0.384 ± 0.061 | −177 ± 130 | 0.802 ± 0.091 | −136 ± 98 | b.d.l. | | 0.070 ± 0.009 |
| LU-32 | 0.848 ± 0.072 | −20 ± 83 | 0.463 ± 0.048 | −8 ± 102 | 0.870 ± 0.076 | −62 ± 81 | 0.437 ± 0.078 | 126 ± 200 | 0.203 ± 0.016 |
| LU-20 | 0.802 ± 0.071 | −73 ± 83 | 0.439 ± 0.042 | −61 ± 89 | 0.769 ± 0.070 | −171 ± 76 | b.d.l. | | 0.111 ± 0.030 |
| LU-41 | 0.648 ± 0.056 | −250 ± 65 | 0.304 ± 0.034 | −349 ± 73 | 0.639 ± 0.058 | −312 ± 63 | 0.141 ± 0.047 | −637 ± 121 | 0.276 ± 0.047 |

[a] $^{182}W/^{184}W$, $^{183}W/^{184}W$, and $^{186}W/^{184}W$ from Jacobsen (2005). $^{179}Hf/^{180}Hf$ and $^{180}Hf/^{184}W$ from Lodders (2010).
[b] $\delta^{i}W/^{184}W$ (‰) = [($^{i}W/^{184}W$)$_{measured}$/($^{i}W/^{184}W$)$_{solar}$ −1] × 10$^3$.
[c] $\delta^{179}Hf/^{180}Hf$ (‰) = [($^{179}Hf/^{180}Hf$)$_{measured}$/($^{179}Hf/^{180}Hf$)$_{solar}$ −1] × 10$^3$.
[d] corrected with relative sensitivity factor determined on the SiC ceramic.
n.a. = not analysed; b.d.l. = below detection limit.

|       | 184 | 185  | 186  | 187  | 188  | 189  |
|-------|-----|------|------|------|------|------|
| Os    | 0.02 | 93.6d | 1.59 | 1.60 | 13.3 | 16.2 |

|       | 185 | 186 | 187 | 188 |
|-------|-----|-----|-----|-----|
| Re    | 37.4 | 3.7d | 42E9a | 17h |

|       | 180  | 181   | 182  | 183  | 184  | 185  | 186  | 187  |
|-------|------|-------|------|------|------|------|------|------|
| W     | 0.12 | 121.2d | 26.5 | 14.3 | 30.6 | 75.1d | 28.4 | 23.7h |

|       | 180 | 181 | 182 | 183 | 184 |
|-------|-----|-----|-----|-----|-----|
| Ta    | 0.01 | 99.99 | 114.4d | 5.1d | 8.7h |

|       | 177 | 178 | 179 | 180 | 181 | 182 | 183 |
|-------|-----|-----|-----|-----|-----|-----|-----|
| Hf    | 18.6 | 27.3 | 13.6 | 35.1 | 42.3d | 9E6a | 1.07h |

s-process

f1.eps

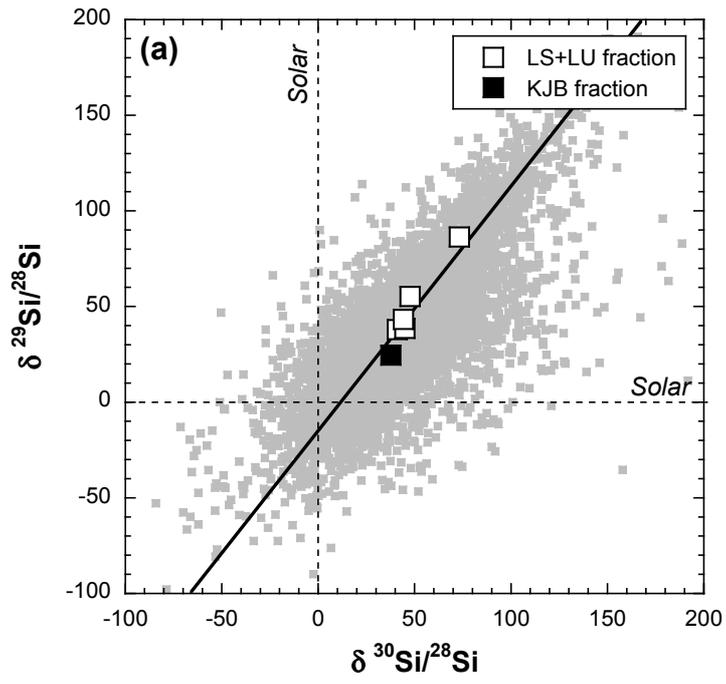

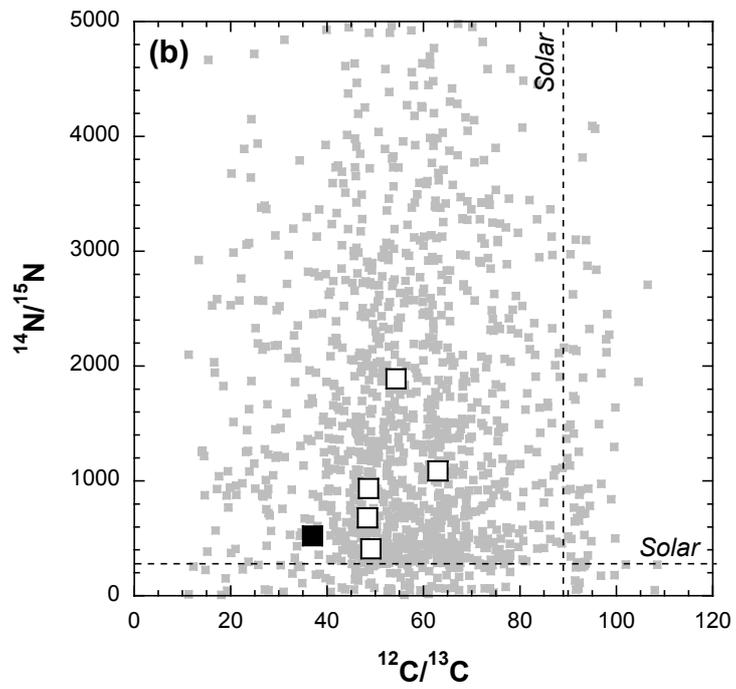

f2.eps

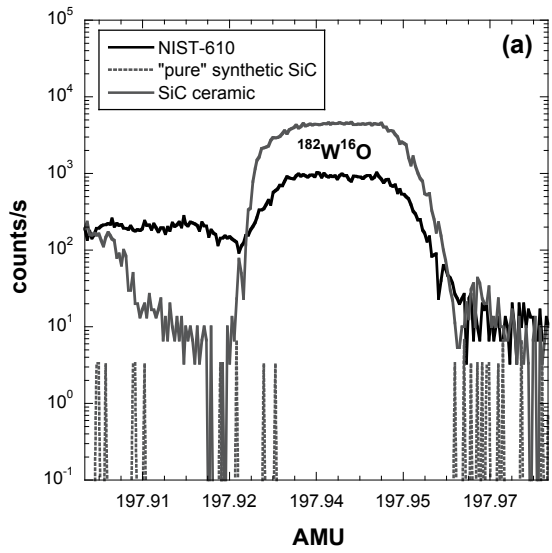
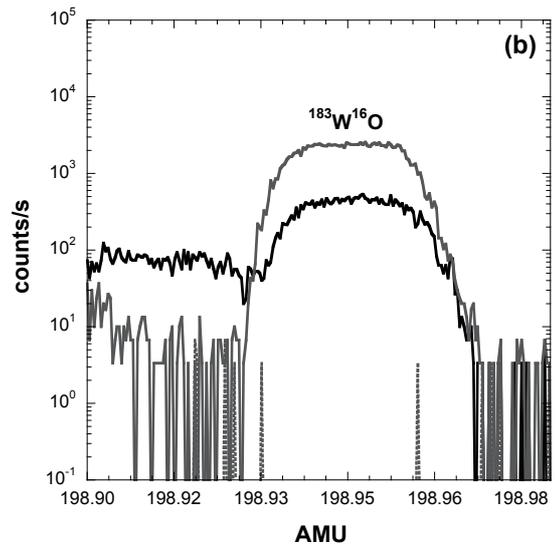
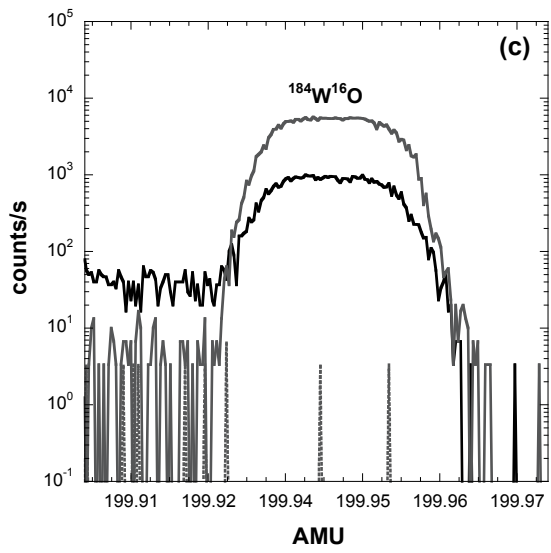
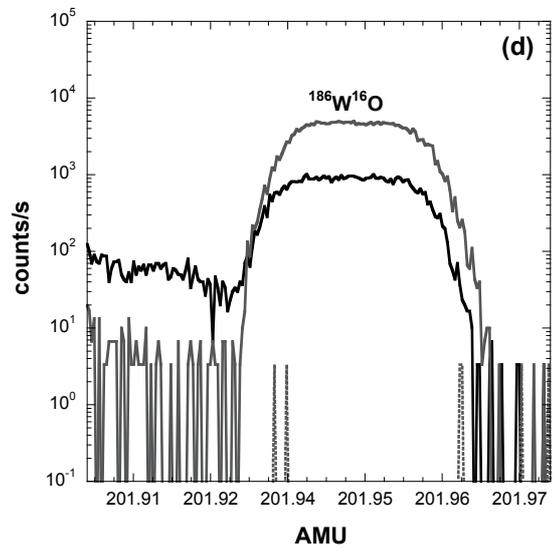

f3.eps

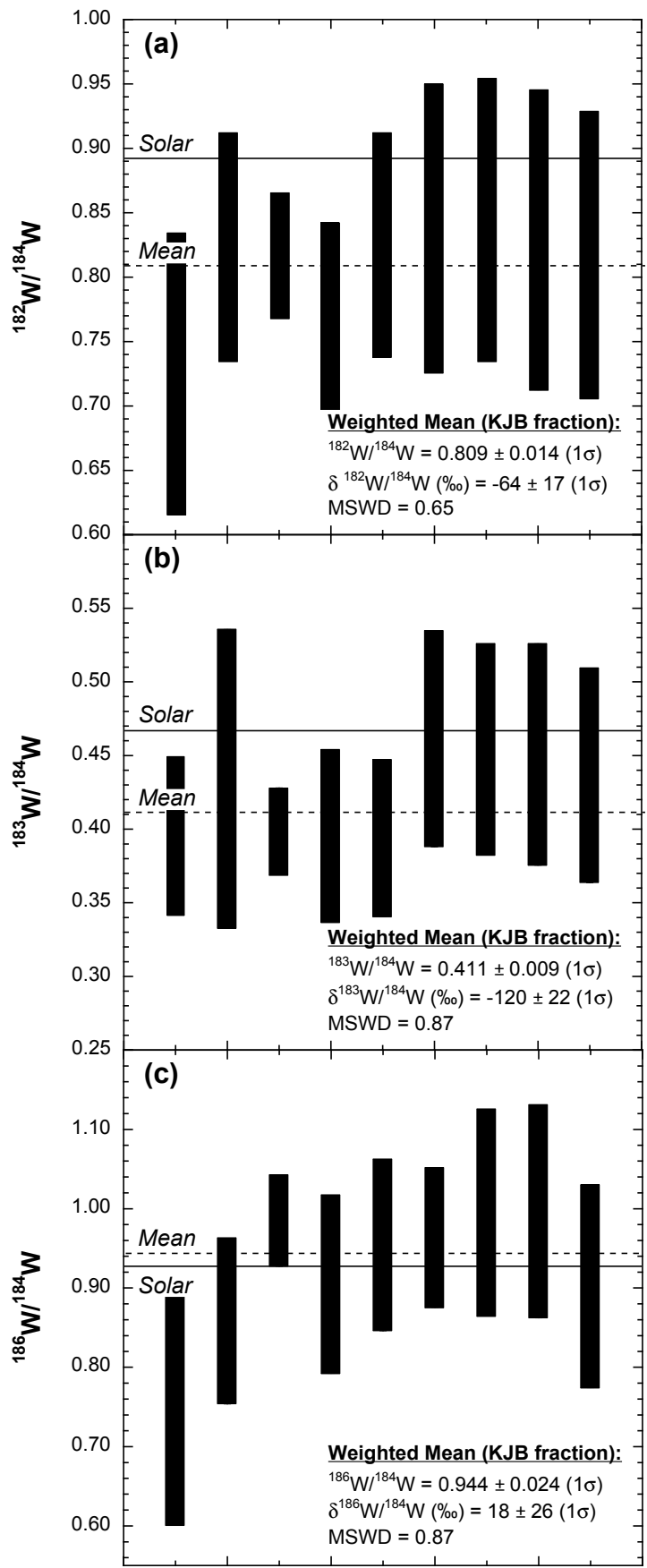
f4.eps

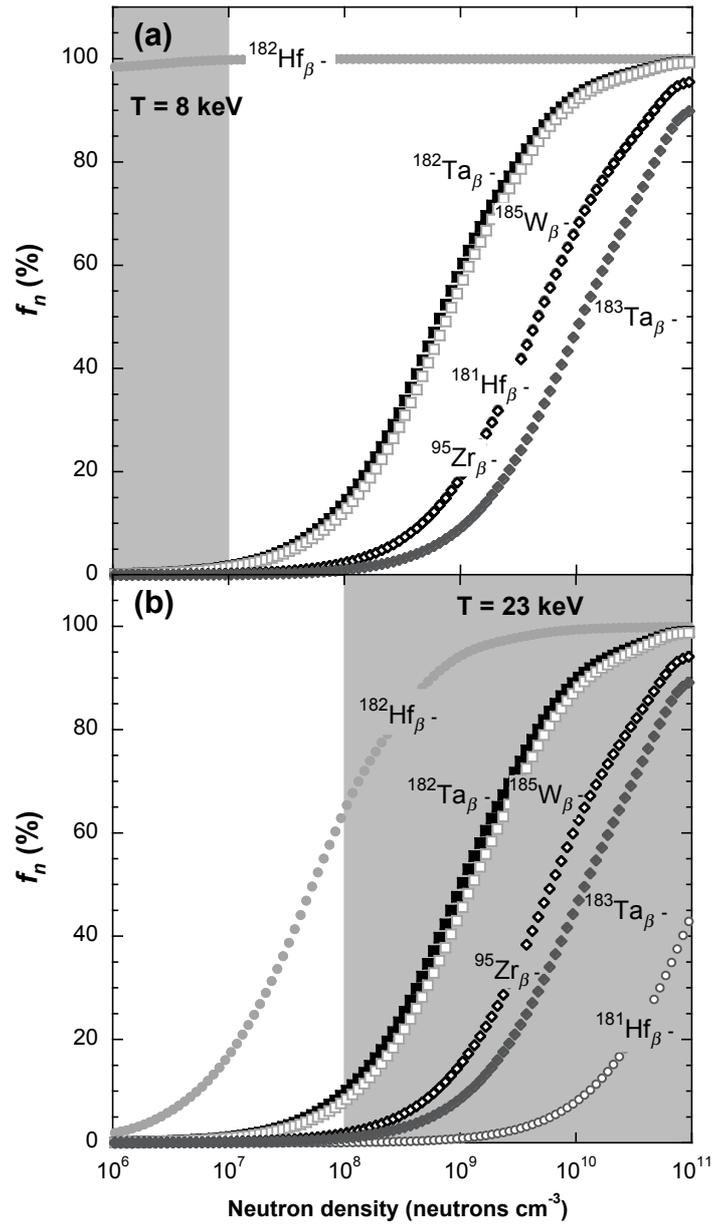

f5.eps

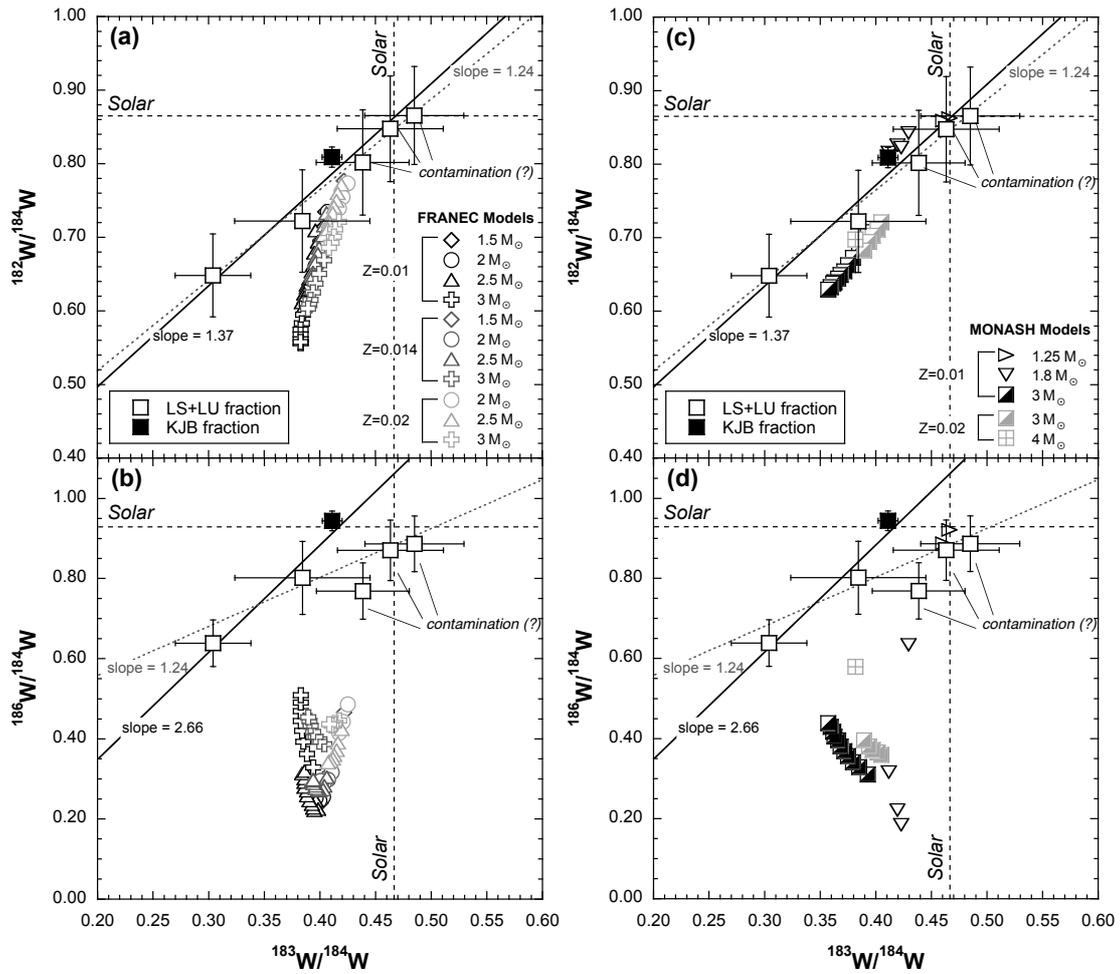

f6.eps

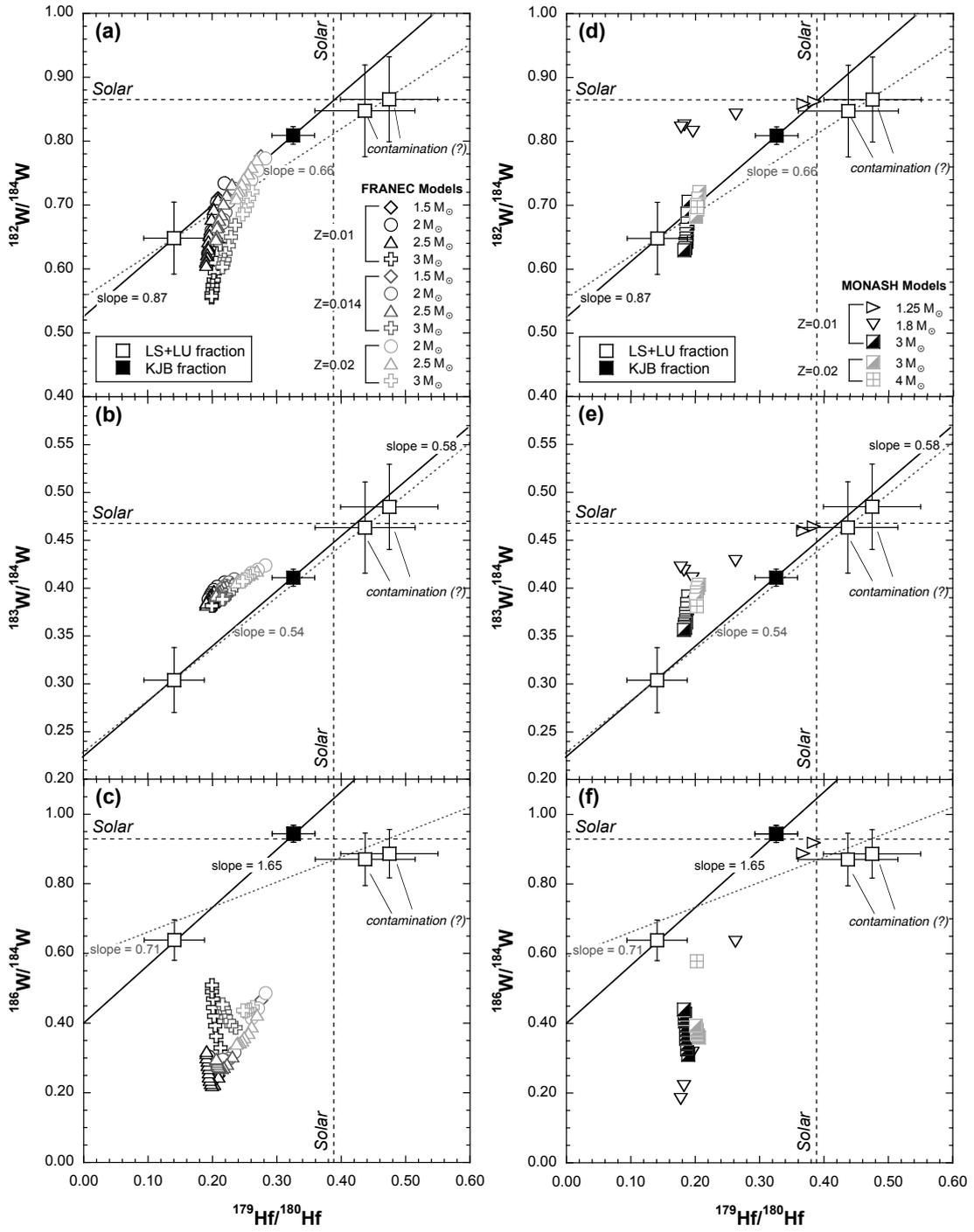

f7.eps